 \documentclass[
reprint,
amsmath,amssymb,
aps,
prl,
]{revtex4-2}

\usepackage{txfonts}

\usepackage{graphicx}
\usepackage{dcolumn}
\usepackage{bm}
\usepackage{xcolor}
\usepackage{ulem}

\begin{document}

\title{
Fractional vortex array realized at twin boundary in a nematic superconductor
}

\author{Sakiko~Noda$^{1,2}$} 
\author{Hiroto~Adachi$^{1,2}$}  
\author{Masanori~Ichioka$^{1,2}$}



\affiliation{
$^1$Research Institute for Interdisciplinary Science, Okayama University, Okayama 700-8530, Japan\\
$^2$Department of Physics, Okayama University, Okayama 700-8530, Japan\\
}

\date{\today}

\begin{abstract}
Within a framework of two-component  Ginzburg-Landau theory as a model of superconducting FeSe, we study the spatial structure of vortex states in the presence of nematic twin boundary in an $s \pm d$ wave nematic superconductor. 
The result shows that the orientation of the nematic vortex core is rotated $90^\circ$ across the twin boundary, and just at the twin boundary the nematic vortex becomes two fractional vortices with the topological nature of core-down and core-up merons. 
The exotic vortex states may be confirmed by observing the time evolution of vortex flow when the vortices are trapped in and escape from the nematic twin boundary. 
\end{abstract}


\maketitle



The superconductivity of a layered iron-chalcogenide FeSe ($T_{\rm c}$=9 K) attracts much attention since it exhibits exotic properties, such as a 
nematic electronic state, in addition to interest of high $T_{\rm c}$ superconductivity in FeSe-based materials~\cite{doi:10.1073/pnas.0807325105,Bohmer_2018,doi:10.7566/JPSJ.89.102002,sym12091402}. 
As for the nematic natures, FeSe exhibits a structural transition at $T_{\rm s} \sim 90$ K, which reduces the tetragonal fourfold symmetry at high temperature $T$ to nematic twofold symmetry at low $T$~\cite{PhysRevB.87.180505}.  
In the nematic phase, there are two domains with nematic orientation along the $a$ or $b$ axis. 
The nematic twin boundary appears parallel to the (110) direction since the twin boundary is a crystallographic mirror plane of two neighboring domains. 
Reflecting the nematic electronic state,~\cite{PhysRevLett.113.237001,PhysRevB.90.121111}
superconducting state at $T<T_{\rm c}$ shows a nematic twofold symmetric superconducting gap structure~\cite{PhysRevX.8.031033,doi:10.1126/science.aal1575}. 
The nematic anisotropy and the nematic twin boundary have been studied by real space observations of electronic local density of states (LDOS) using scanning tunneling microscopy (STM)~\cite{doi:10.1126/science.1202226, PhysRevLett.109.137004, PhysRevX.5.031022, PhysRevB.53.2835, NatComm.9-282, npjQM.3-12, doi:10.1021/acs.nanolett.3c00125}. 
There, when magnetic fields penetrate superconductors as a vortex with a flux quantum, the vortex shows a twofold symmetric vortex core image reflecting the nematic orientation, and some vortices are trapped at the nematic twin boundary. 
Theoretically, the nematic superconducting gap is expressed as $s \pm d$ wave pairing state by the combination of $s$ and $d$ wave order parameters. 
Accordingly, the two-component Ginzburg-Landau (GL) theory is used in phenomenological studies about the spatial structure of superconducting states, such as the nematic vortex state~\cite{PhysRevB.99.144514, npjQM.3-12} or the nematic twin boundary~\cite{PhysRevX.5.031022, PhysRevB.53.2835} in ${\rm FeSe}$. 
Thus, we need a theoretical study also about vortex structure just at the twin boundary. 

While conventional vortices have a flux quantum with phase winding $2 \pi$ of superconducting order parameter around a vortex center, half quantum vortices can appear when a $\pi$ Josephson junction with $\pi$-phase shift exists in the phase winding around a vortex~\cite{PismaZhETF.25.314,PhysRevLett.92.257005,PhysRevLett.101.247001}.  
This situation occurs at the tri-crystal grain boundary of a $d$ wave 
high $T_{\rm c}$ cuprate superconductor, or the corner junction between $d$ wave and $s$ wave superconductors~\cite{Kirtley_2010,PhysRevLett.73.593,nature01442}.  
At the bi-crystal grain boundary, a high density of facets can make splintered Josephson vortices with a fractional flux quantum of less than half~\cite{PhysRevLett.77.2782}.  
These half quantum or fractional vortices appear at a zero field, and the vortex position is fixed by the corner or facet of the grain boundary.
Another type of half quantum vortices or fractional vortices are predicted in multi-component superfluids~\cite{Nature324-333-1986,doi:10.1073/pnas.96.14.7760,PhysRevLett.127.095302,PhysRevA.107.053304} and superconductors~\cite{RevModPhys.63.239,10.1143/PTP.102.965,PhysRevLett.89.067001,PhysRevLett.92.157001,PhysRevB.70.100502,PhysRevB.71.172510,TANAKA201844,PhysRevLett.119.167001,PhysRevResearch.2.043192,PhysRevB.83.144519,PhysRevB.86.024512}.
Since the fractional vortex is an interesting topological object related to domain walls or kinks realized by the combination of multi-component order parameters, it has been searched for a long time. 
However, the experimental detection is rare 
in the multi-component superconductors.
Recently possible fractional vortex was reported in one of Fe-based superconductors~\cite{doi:10.1126/science.abp9979}. 
Therefore it is desirable to study the possibility of  fractional vortex also in the case of multi-component $s \pm d$ wave nematic superconductor FeSe.  

The purpose of this study is  to clarify vortex states of $s \pm d$ wave superconductor in the presence of nematic twin boundary, by numerical simulation of time-dependent Ginzburg-Landau (TDGL) theory. 
We examine the spatial structure  of fractional vortices appearing at the nematic twin boundary, within a framework of two-component GL theory suggested as a model of superconducting FeSe. 
We also study vortex flow when current is applied parallel to the nematic twin boundary, to see the dynamical process of how vortices are trapped to the twin boundary.


The formulation of our study is as follows.
The nematic twofold symmetric superconducting gap structure observed in FeSe ~\cite{PhysRevX.8.031033,doi:10.1126/science.aal1575} is described as  $s \pm d$ wave pairing~\cite{doi:10.7566/JPSJ.89.102002,PhysRevX.5.031022} 
given by 
\begin{eqnarray}
\Delta({\bm r},{\bm k})=\Delta_s({\bm r})+\Delta_d({\bm r})\phi_d({\bm k})
\end{eqnarray}
with $\phi_d({\bm k})=\hat{k}_x^2 -\hat{k}_y^2$, 
assuming   $\Delta_d \sim \pm \Delta_s$ in the uniform $s \pm d$ wave pairing state. 
There, the gap minimum is located at the direction of $k_x=0$ ($k_y=0$) in the momentum space for $s+d$ ($s-d$) wave pairing. 
Near the nematic twin boundary between $s+d$ wave and $s-d$ wave superconducting regions, two order parameters of $s+d$ wave and $s-d$ wave are mixed by the proximity effect penetrating to the other region. 
Therefore, the spatial structure of the mixed local superconducting state is described by the spatial variation of two component $s$ wave and $d$ wave order parameter, instead of $s+d$ wave and $s-d$ wave in this study. 

Within the GL approximation 
for  the two-component order parameters $\Delta_s({\bm r})$ and $\Delta_d({\bm r})$, 
the free energy in the superconducting state is generally given by
\begin{eqnarray} 
F_s=
\int \{ f({\bm r}) + f_{\rm B}({\bm r}) \}
{\it d}{\bm r} 
\end{eqnarray} 
with
\begin{eqnarray} && 
f({\bm r})= - ( 1 - T) |\Delta_s|^2 + a_d ( 1 - T/T_{{\rm c}d}) |\Delta_d|^2 
\nonumber \\ && \quad 
+ c \epsilon (\Delta_s^\ast \Delta_d + \Delta_d^\ast \Delta_s) 
+ \frac{1}{2} \{ |\Delta_s|^4 + b_d|\Delta_d|^4 
\nonumber \\ && \quad 
+ \gamma_1 |\Delta_s|^2 |\Delta_d|^2  
        +\frac{\gamma_2}{2}  (\Delta_s^{\ast 2} \Delta_d^2 + \Delta_d^{\ast 2} \Delta_s^2 ) \}
\nonumber \\ && \quad 
+\Delta_s^\ast (\Pi_x^2 + \Pi_y^2 )\Delta_s 
+K_d \Delta_d^\ast (\Pi_x^2 + \Pi_y^2 )\Delta_d
\nonumber \\ && \quad 
 +\frac{\tilde K}{2}  \{ 
  \Delta_s^\ast (\Pi_x^2 - \Pi_y^2 )\Delta_d 
+\Delta_d^\ast (\Pi_x^2 - \Pi_y^2 )\Delta_s  \}
\label{eq:GL-fe}
\end{eqnarray} 
and $f_{\rm B}=\kappa^2 |{\bm B}({\bm r})-{\bm H}|^2$ 
in dimension-less form~\cite{npjQM.3-12,PhysRevX.5.031022, PhysRevB.53.2835}. 
$\Pi_x = -
{\it i}\partial_x +A_x$ and $\Pi_y = -
{\it i}\partial_y +A_y$
with vector potential ${\bm A}=(A_x,A_y,0)$.
In Eq. (\ref{eq:GL-fe}), to stabilize $s \pm d$ wave pairing superconductivity observed in FeSe, we introduce the Josephson coupling term with a factor $c \epsilon$ as in previous studies~\cite{PhysRevB.99.144514, npjQM.3-12,PhysRevX.5.031022, PhysRevB.53.2835}, which comes from the coupling of $s$ wave and $d$ wave superconducting order parameters 
by the orthorhombic distortion $\epsilon \equiv \epsilon_{xx} -\epsilon_{yy}$~\cite{PhysRevX.5.031022, PhysRevB.53.2835} in the nematic electronic states reflecting an imbalance of Fe $d_{xz}$ and $d_{yz}$ orbitals~\cite{PhysRevLett.113.237001,PhysRevB.90.121111}. 
In our calculation, we set $a_d=b_d=K_d=T_{{\rm c}d}=1$, $|c \epsilon|=1.0$, $\gamma_1=1.2$, $\gamma_2=1.0$, ${\tilde K}=1/\sqrt{2}$, GL parameter $\kappa=5$, and $T=0.1$. 

The vortex structure is numerically calculated by the time evolution following the TDGL equation coupled with the Maxwell equation~\cite{PhysRevB.47.8016, PhysRevLett.71.3206, PhysRevLett.92.157001, PhysRevB.70.100502, PhysRevB.71.172510, SADOVSKYY2015639}, 
\begin{eqnarray} && 
\left( 
\frac{\partial}{\partial t} 
- 
{\it i}\phi 
\right)
\Delta_j({\bm r})= -\frac{1}{12}\frac{\partial f}{\partial \Delta^\ast_j({\bm r})}, \quad (j = s,d)
\\ && 
\frac{\partial}{\partial t}{\bm A}({\bm r}) +\nabla\phi
= {\bm J}({\bm r}) - \kappa^2 \nabla \times {\bm B}({\bm r}), 
\end{eqnarray}
with ${\bm B}({\bm r})=\nabla \times{\bm A}({\bm r})$ and the supercurrent 
\begin{eqnarray} 
{\bm J}({\bm r}) =
(J_x,J_y,J_z)
=-\left( 
\frac{\partial f}{\partial A_x({\bm r})}, \frac{\partial f}{\partial A_y({\bm r})},0
\right)  . 
\label{eq:Jsuper}
\end{eqnarray}
About the scalar potential $\phi$, for simplicity we use the gauge fixing condition of $\phi =0$ in our calculation~\cite{PhysRevLett.71.3206,SADOVSKYY2015639}.  

The original derivation of the TDGL equation was done in gapless dirty $s$ wave superconductor~\cite{Schmid1966} to consider dissipation by gapless excitation within the full gap of $s$ wave pairing for relaxation of the order parameter. 
However, simulations of the TDGL equation are now satisfactorily used to study vortex dynamics of $d$ wave superconductors not in the dirty limit, including high $T_{\rm c}$ cuprate superconductors~\cite{SADOVSKYY2015639,PhysRevB.93.060508},
which have gapless excitation due to nodes of the superconducting gap. 

\begin{figure}[tb]
\begin{center}
\includegraphics[width=8.3cm]{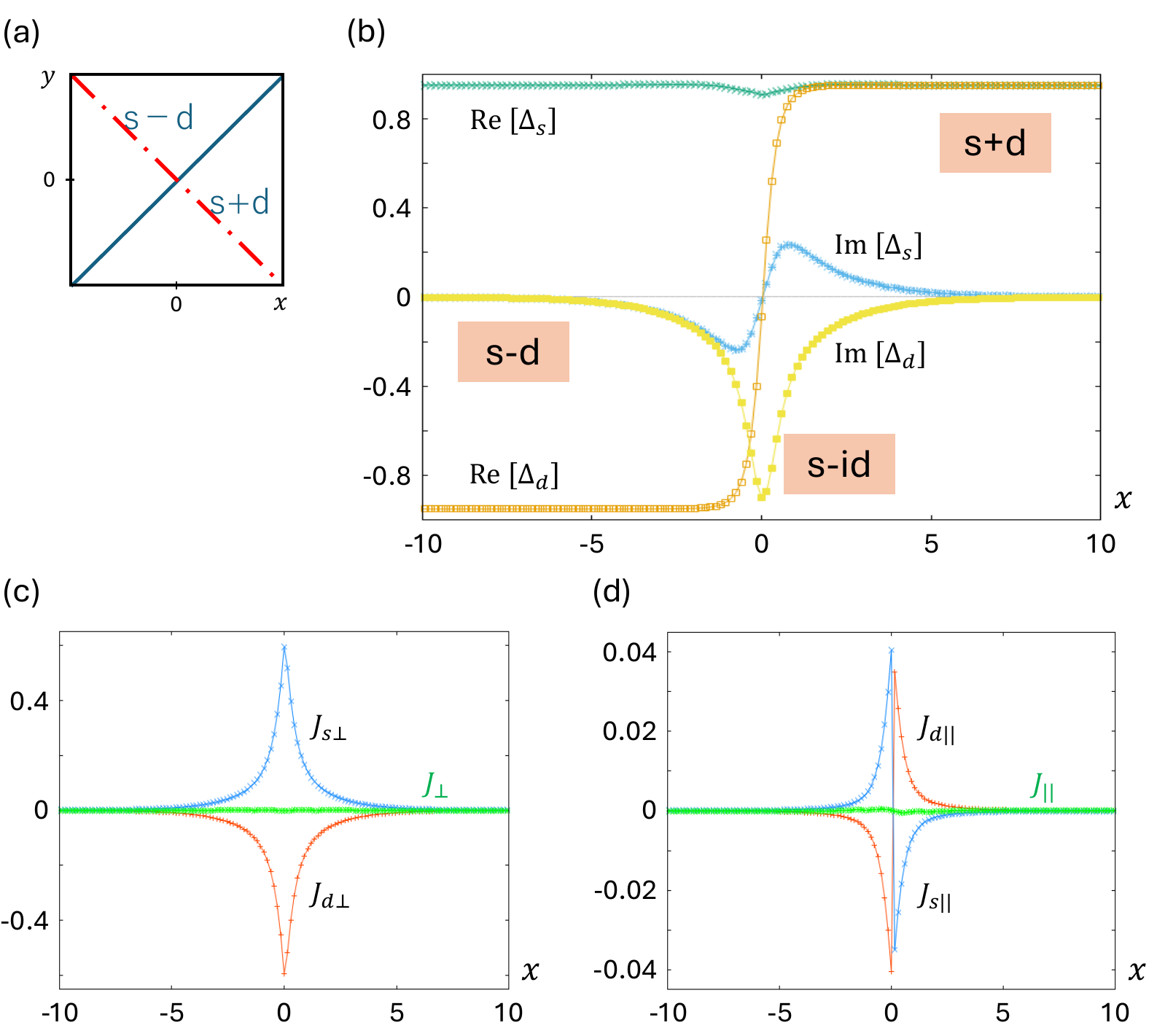}
\end{center}
\caption{\label{fig:B0}
(Color online) 
(a) 
The square region of our calculation is schematically presented. 
The twin boundary is located at a diagonal solid line $y=x$, 
and the upper left (lower right) region is assigned to the $s-d$ ($s+d$) wave pairing domain. 
(b) 
Spatial variation of order parameters at a zero field $H=0$. 
We show ${\rm Re}\Delta_s({\bm r})$,  ${\rm Im}\Delta_s({\bm r})$,  
${\rm Re}\Delta_d({\bm r})$, and ${\rm Im}\Delta_d({\bm r})$  
as a function of $x$ along a dash-dot line $y=-x$ in (a).
(c) Perpendicular supercurrent component $J_\perp=J_{s\perp}+J_{d\perp}+J_{sd\perp}$ and (d) parallel component $J_\parallel=J_{s\parallel}+J_{d\parallel}+J_{sd\parallel}$ to the nematic twin boundary along the same line as in (b). 
Here, $J_s$, $J_{d}$, and $J_{sd}$ are defined in Eqs. (\ref{eq:Js}), (\ref{eq:Jd}), and (\ref{eq:Jsd}). 
Bold lines show $J_\perp = 0$ in (c) and small $J_\parallel$ in (d). 
$J_{sd\perp}= J_{sd\parallel}= 0$ within numerical accuracy. 
}
\end{figure}

\begin{figure}
\begin{center}
\includegraphics[width=8.3cm]{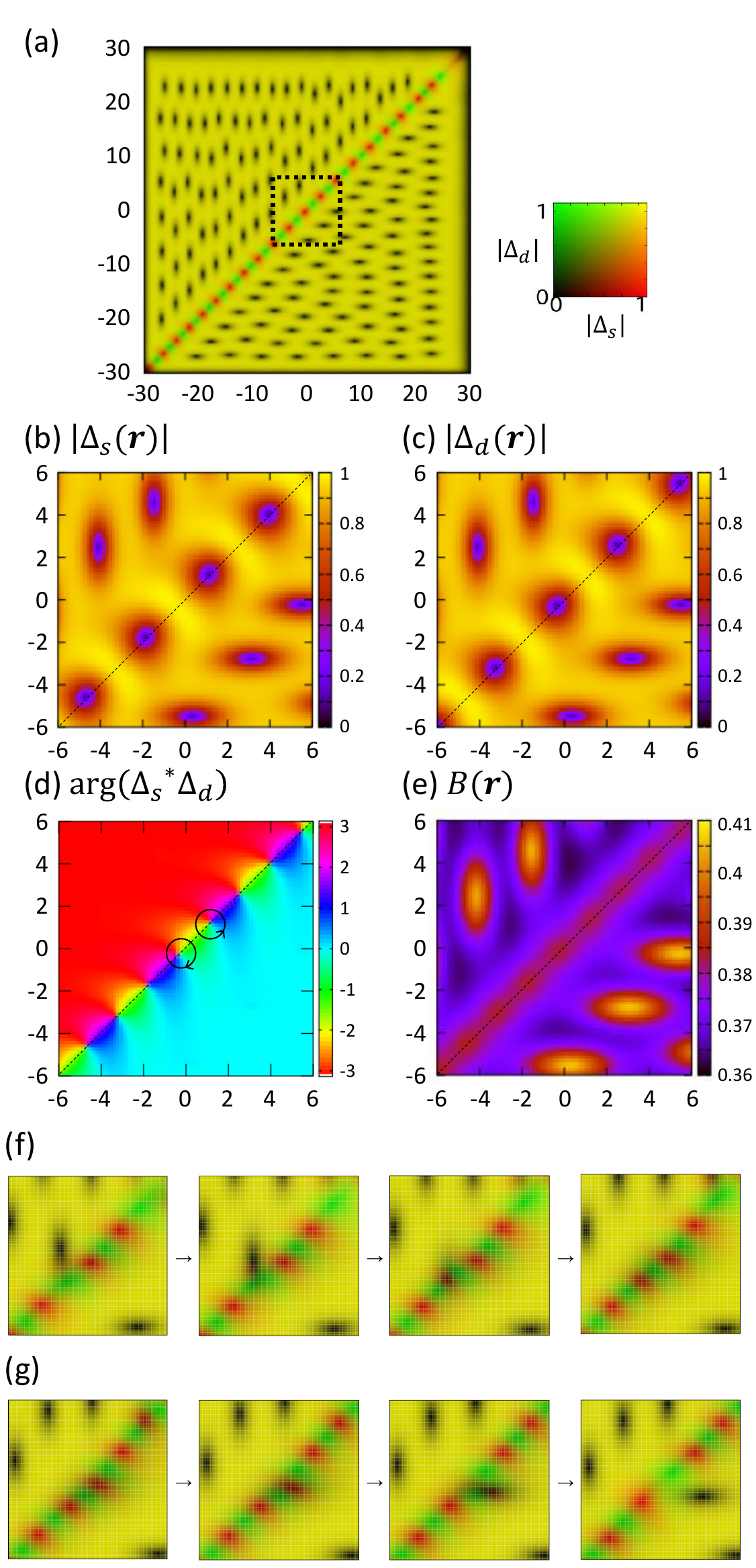}
\end{center}
\caption{\label{fig:B4w}
(Color) 
Spatial variation of the vortex states at $H=0.4$. 
The nematic twin boundary is located at $y=x$. 
(a) 
Color plot of order parameters $|\Delta_s({\bm r})|$ and $|\Delta_d({\bm r})|$ in whole region of the calculation.  
Black cores show conventional vortices. Red and green cores are for fractional vortices. 
(b)  
$s$ wave order parameter $|\Delta_s({\bm r})|$.  
(c) 
$d$ wave order parameter $|\Delta_d({\bm r})|$.  
(d) 
Relative phase ${\rm arg}\{  \Delta_s^\ast({\bm r}) \Delta_d({\bm r}) \}$ 
between $s$ and $d$ wave order parameters. Circular arrows present the phase winding at a fractional vortex. 
(e) 
Internal magnetic field $B({\bm r})$.
In (b)-(e), the square region of dashed lines in (a) is enlarged, and diagonal lines show the locations of nematic twin boundaries.  
(f) and (g) 
Snapshots in the time-evolution of vortex flow when supercurrent is applied parallel to the nematic twin boundary.
The color scale for $|\Delta_s({\bm r})|$ and $|\Delta_d({\bm r})|$ is the same as in (a).   
In (f), a conventional vortex is trapped to the twin boundary and changes to two fractional vortices. 
In (g), the fractional vortices escape from the twin boundary. 
}
\end{figure}

We perform numerical calculations in a square region of size $60 \times 60$ in units of coherence length, as shown in Figs. \ref{fig:B0}(a) and \ref{fig:B4w}(a). 
Outside of the region, we assume $\Delta_s({\bm r})= \Delta_d({\bm r})=0$ and ${\bm B}({\bm r}) = {\bm H}$ with applied magnetic field ${\bm H}=(0,0,H)$.  
The nematic twin boundary is assumed to be on a diagonal line $y=x$. 
We set $c \epsilon =1$ in the $s-d$ wave pairing domain at $y > x$, 
and $c \epsilon =-1$ in the $s+d$ wave pairing domain at $y < x$~\cite{PhysRevX.5.031022}. 
The barrier potential at the twin-boundary is not considered, since effects of charging, stress and
mismatch of atomic lattice are not significant at the nematic twin boundary in FeSe as observed by STM~\cite{PhysRevX.5.031022}. 
Therefore, we can study structures of the vortex and nematic twin boundary in superconducting FeSe by a simple ideal model of the two-component GL theory. 



\begin{figure}[h]
\begin{center}
\includegraphics[width=8.3cm]{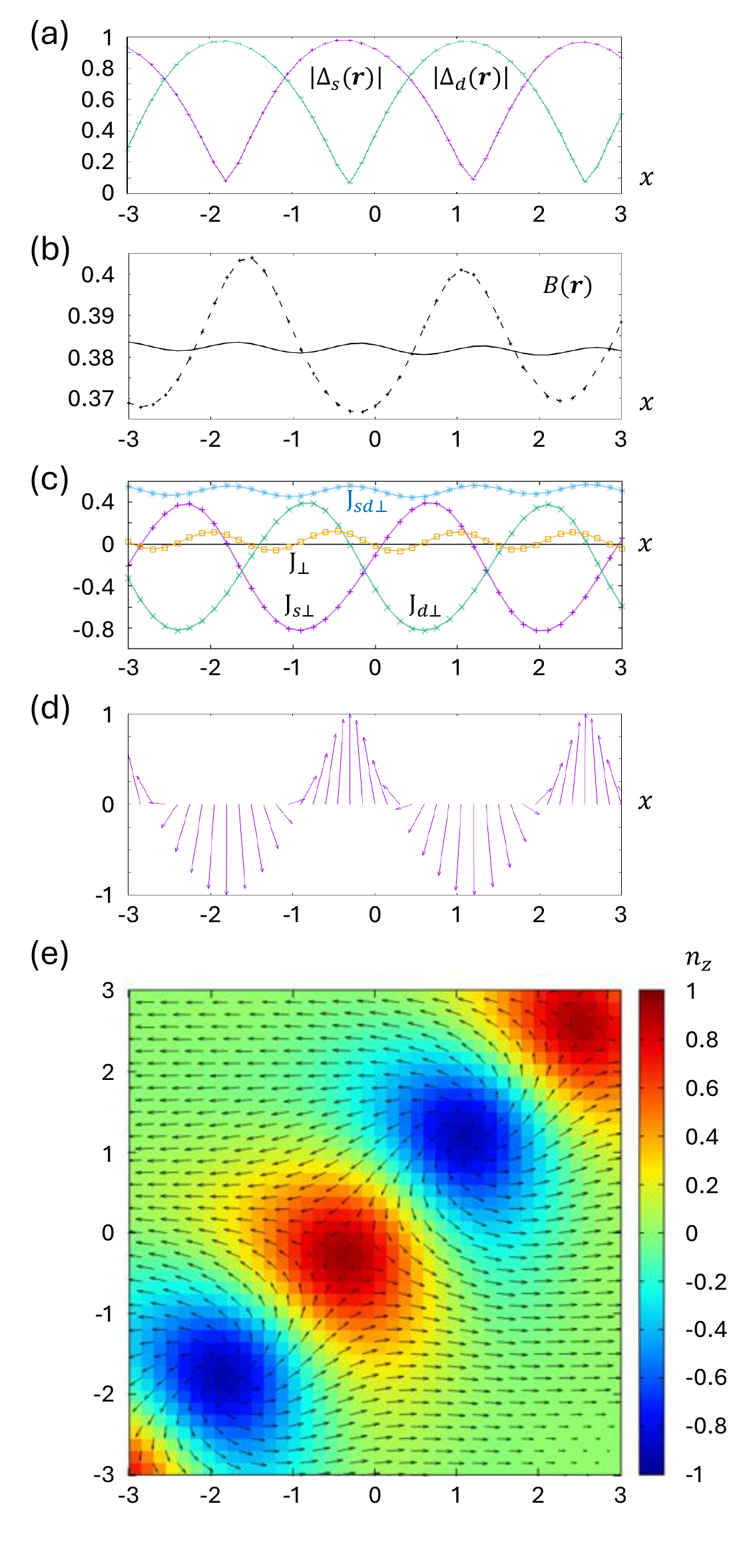}
\end{center}
\caption{\label{fig:B4}
(Color) 
Profile of 
(a) $|\Delta_s({\bm r})|$, $|\Delta_d({\bm r})|$,  
(b) $B({\bm r})$ (solid line), 
(c) $J_\perp({\bm r})=J_{s\perp}({\bm r})+J_{d\perp}({\bm r})+J_{sd\perp}({\bm r})$, 
and (d) ${\bm n}({\bm r})$ 
as a function of $x$ along the nematic twin boundary $y=x$. 
$H=0.4$. 
In (b) the dashed line shows $B({\bm r})$ along line $y=x+6.3$ for conventional vortices. 
(e) Vector ${\bm n}({\bm r})$ in the focused region $6 \times 6$ near the twin boundary, where arrows show vector $(n_x,n_y)$, and $n_z$ is presented by color. 
}
\end{figure}
Before studying the vortex states, we see spatial variation of order parameters near a nematic twin boundary at a zero field $H=0$ in Fig. \ref{fig:B0}(b). 
There, ${\rm Re}\Delta_d({\bm r})$ changes the sign from negative ($s-d$ wave) to positive ($s+d$ wave) across the twin boundary at $x=0$, while ${\rm Re}\Delta_s({\bm r})$ keeps almost constant.  
As suggested in Refs. \cite{PhysRevX.5.031022} and \cite{PhysRevB.53.2835}, 
non-zero ${\rm Im}\Delta_d({\bm r})$ appears near the twin boundary, indicating that time reversal symmetry is locally broken by the 
$s-{\it i}d$ wave, and the amplitude $|\Delta_d({\bm r})|$ does not vanish there. 
We also find finite ${\rm Im}\Delta_s({\bm r})$ which was not considered in the previous analytic consideration~\cite{PhysRevX.5.031022, PhysRevB.53.2835}. 
Since we consider a straight line of nematic twin boundary without facets and corners, fractional vortices do not appear at a zero field. 
Therefore, the fractional vortex states appearing in the $s \pm d$ wave superconductor under magnetic fields which we discuss later are different from those appearing at grain boundary in {\it single} component $d$ wave superconductors~\cite{Kirtley_2010,PhysRevLett.73.593,nature01442,PhysRevLett.77.2782}. 

To consider internal current by nonuniform phase of the order parameters 
at the zero field, we decompose the supercurrent in Eq. (\ref{eq:Jsuper}) to three parts as ${\bm J}=(J_x,J_y)={\bm J}_s + {\bm J}_d +{\bm J}_{sd}$ with 
\begin{eqnarray} && 
{\bm J}_s = -{\rm Re}\left\{  \Delta^\ast_s (\Pi_x,\Pi_y) \Delta_s \right\} , 
\quad 
\label{eq:Js}
\\  && 
{\bm J}_d = -
K_d {\rm Re}\left\{  \Delta^\ast_d (\Pi_x,\Pi_y) \Delta_d \right\} , 
\quad
\label{eq:Jd} 
\\ && 
{\bm J}_{sd} = -
\frac{\tilde K}{2} 
{\rm Re}\left\{  \Delta^\ast_s (\Pi_x,-\Pi_y) \Delta_d  
                                +\Delta^\ast_d (\Pi_x,-\Pi_y) \Delta_s   \right\} . 
\label{eq:Jsd} \qquad 
\end{eqnarray}
${\bm J}_s$ and  ${\bm J}_d$ are, respectively, components from order parameters $\Delta_s$ and $\Delta_d$. 
${\bm J}_{sd}$ comes from cross terms of $\Delta_s$ and $\Delta_d$. 
In Fig. \ref{fig:B0}(c), we show perpendicular component $J_\perp=(J_x-J_y)/\sqrt{2}$ of supercurrent to the twin-boundary, with the decomposition of $J_\perp=J_{s\perp}+J_{d\perp}+J_{sd\perp}$. 
There, components $J_{s\perp}$ and $J_{d\perp}$ have finite values near the nematic twin boundary, due to the nonuniform phase of order parameters $\Delta_s$ and $\Delta_d$ and finite vector potential ${\bm A}$ appearing near the twin boundary. 
However, since $J_{s\perp}$ and $J_{d\perp}$ cancel each other and $J_{sd\perp} \sim 0$, the net current $J_\perp $ perpendicular to the twin-boundary vanishes at a zero field within the numerical accuracy of the calculation, as expected by the current conservation $\nabla\cdot {\bm J}=0$ and the symmetry of the system.  
In Fig. \ref{fig:B0}(d), we show parallel component $J_\parallel=(J_x+J_y)/\sqrt{2}$ of supercurrent to the twin-boundary, with the decomposition of $J_\parallel=J_{s\parallel}+J_{d\parallel}+J_{sd\parallel}$. 
There, $J_{s\parallel}$ and $J_{d\parallel}$ flow in the opposite directions without a perfect cancellation, and $J_{sd\parallel} \sim 0$. 
Thus, near the twin boundary, a small supercurrent parallel to the boundary remains.


Next, we study the vortex states under a magnetic field. 
Figure \ref{fig:B4w}(a) presents the spatial variation of $|\Delta_s({\bm r})|$ and $|\Delta_d({\bm r})|$ at $H=0.4$ in the whole region of the calculation. 
There, the dark region of the vortex core has an elliptic shape extending to the $x$ direction, reflecting the nemacity  in the $s+d$ wave domain at $y < x$. 
The nematic anisotropy comes from contributions of $\tilde{K}$-term in Eq. (\ref{eq:GL-fe})~\cite{npjQM.3-12}.  
On the other hand, in the $s-d$ wave domain at $y > x$ the nematic vortex core is extending to the $y$ direction. 
The change of nemacity of the vortex core across the twin boundary is seen in the LDOS observed by STM~\cite{PhysRevLett.109.137004, PhysRevX.5.031022}, 
while it is noted that the vortex core shape of the order parameter does not necessarily coincide with that of the LDOS~\cite{sym12010175}.

Vortices on the twin boundary are presented in red or green colors in Fig. \ref{fig:B4w}(a). 
To see the vortex states in detail, we focus on the region near the twin boundary in Fig. \ref{fig:B4w}(b)-(e). 
Comparing (b) $|\Delta_s({\bm r})|$ and (c) $|\Delta_d({\bm r})|$, away from the twin boundary, both components have vortex core at the same position. 
However, the vortex core positions are separated from each other on the twin boundary at $y=x$. 
There, at a vortex of red color in Fig. \ref{fig:B4w}(a) $|\Delta_d({\bm r})|$ has a vortex core where $|\Delta_s({\bm r})|$ has its maximum, and vice versa at a vortex of green color. 
These structures are also seen in Fig. \ref{fig:B4}(a).  
The spatial variation of relative phase ${\rm arg}\{  \Delta_s^\ast({\bm r}) \Delta_d({\bm r}) \}$ 
between $s$ and $d$ wave order parameters is presented in Fig. \ref{fig:B4w}(d).  
Away from the twin boundary, the relative phase is locked to be near 0 ($\pm \pi$) in the domain of $s+d$ ($s-d$) wave state at $y<x$ ($y>x$), including the vortex core region. 
On the twin boundary, the relative phase becomes $\pi/2$ of 
$s+{\it i}d$ or $-\pi/2$ of 
$s-{\it i}d$. 
This helps that the relative phase has winding $2\pi$ ($-2 \pi$) around a vortex of order parameter $\Delta_d$ ($\Delta_s$) on the twin boundary. 
Therefore, vortices on the twin boundary become fractional vortices by separating vortex cores of $s$ and $d$ wave order parameters. 
Since both order parameters give almost equal weight to the contribution in the present choice of parameters, the fractional vortices can be treated as half-quantum vortices.
The internal magnetic field $B({\bm r})$ around vortices is shown in Figs. \ref{fig:B4w}(e) and \ref{fig:B4}(b).  
There, the height of $B({\bm r})$ is lower on the twin boundary, compared with the conventional vortex away from the twin boundary.
This is because screening current around a half-quantum vortex comes from only the $s$ or $d$ wave order parameter, and the magnetic flux of a vortex becomes half.
We note that fine-tuning of parameters is necessary for quantitative comparison with experimental observation~\cite{PhysRevB.100.024514}. 
Perpendicular supercurrent component $J_\perp({\bm r})$  crossing the twin boundary is shown in Fig. \ref{fig:B4}(c) with the decomposition $J_\perp=J_{s\perp}+J_{d\perp}+J_{sd\perp}$. 
There, $J_{s\perp}$ and $J_{d\perp}$ reflect the circular screening current around vortices of $\Delta_s({\bm r})$ and $\Delta_d({\bm r})$, respectively. 
And $J_{sd\perp}$ gives finite contribution. 
Combining these contributions, $J_\perp$ gives screening current around fractional vortices.   
The spatial average of $J_\perp$ shows 
a slight shift from zero, reflecting that the current has a small component of diamagnetic response to the magnetic field penetrating from the outer boundary, in addition to the screening current around vortices. This small effect survives even at the center region since the effective penetration lengths at the upper-right and lower-left sides are different, as shown in Fig. \ref{fig:B4w}(a).

We also perform some additional calculations.  
When parameters are changed to $a_d=0.8$ or $T_{{\rm c}d}=0.8$ so that $s$ and $d$ wave components are not symmetric, we see similar fractional vortex array structure, while a maximum of $|\Delta_s({\bm r})|$ and $|\Delta_d({\bm r})|$ shows a difference.  
Even when the twin boundary runs along the $x$ direction, similar fractional vortices appear.


To discuss the topological nature of the fractional vortex, we introduce the unit vector ${\bm n}=(n_x,n_y,n_z)$ defined as~\cite{PhysRevLett.119.167001} 
\begin{eqnarray}
n_i =  \frac{ \eta^\dagger \hat{\sigma}_i \eta }{ \eta^\dagger \eta }, \quad 
(i=x,y,z)  
\end{eqnarray} 
with Pauli matrix $\hat\sigma_i$ and $\eta^\dagger =(\Delta_s^\ast, \Delta_d^\ast )$. 
$ \eta^\dagger \eta  = |\Delta_s({\bm r})|^2 + |\Delta_d({\bm r})|^2 $. 
Along the nematic twin boundary, as shown in Fig. \ref{fig:B4}(d), the $z$ component $n_z =( |\Delta_s({\bm r})|^2 - |\Delta_d({\bm r})|)/\eta^\dagger \eta$ changes from $-1$ at the vortex center of $|\Delta_s({\bm r})|=0$, to $+1$ at the vortex center of $|\Delta_d({\bm r})|=0$.    
The spatial variation of ${\bm n}$ is shown in Fig. \ref{fig:B4}(e).  
Orientation of in-plane component $(n_x,n_y)$ reflects the relative phase shown in Fig. \ref{fig:B4w}(d), since   $n_x =2 {\rm Re}\{ \Delta_s^\ast({\bm r}) \Delta_d({\bm r}) \}  /\eta^\dagger \eta$ and $n_y =2 {\rm Im}\{ \Delta_s^\ast({\bm r}) \Delta_d({\bm r}) \}  /\eta^\dagger \eta$. 
Therefore, $(n_x,n_y) \propto (1,0)$ in the $s+d$ wave domain at $y<x$ including cores of conventional vortices, and $(n_x,n_y) \propto (-1,0)$ in the $s-d$ wave domain at $y>x$. 
Along the nematic twin boundary between fractional vortices, $(n_x,n_y) \propto (0,\pm1)$ at the region of 
$s \pm{\it i}d$ wave  with the relative phase $\pm\pi/2$. 
Thus, the orientation of $(n_x,n_y)$ rotates by vorticity $+1$ ($-1$) around a vortex of $\Delta_s$ ($\Delta_d$).
The vortex of $\Delta_d$ ($\Delta_s$) is a core-up (core-down) meron structure with $0 \le n_z \le 1$ ($-1 \le n_z \le 0$)~\cite{Nature564-95-2018}. 
They are connected by the nematic twin boundary. 
The skyrmion number $Q$ is given by  
\begin{eqnarray}
    Q=\frac{1}{4\pi}\int {\bm n} \cdot 
    (\partial_x{\bm n} \times \partial_y{\bm n}) \ 
    {\it d}x \ {\it d}y , 
\end{eqnarray}
The integral region around a fractional vortex in Fig.  \ref{fig:B4}(e) gives the same sign as $Q= -\frac{1}{2}$ for each core-up and core-down meron~\cite{Nature564-95-2018}.  
Thus, the region of the fractional vortex pair of core-up and core-down merons gives a skyrmion of $Q=-\frac{1}{2}-\frac{1}{2}=-1$. 
This is a similar situation to that of the fractional vortex pair in multi-component superconductors~\cite{PhysRevLett.119.167001} and two-dimensional magnets~\cite{Nature564-95-2018,PhysRevB.91.224407, NatComm12.185.2021}. 
Along the nematic twin boundary, since core-up and core-down merons are arranged alternately, we see the structure of the skyrmion lattice as shown in Fig. \ref{fig:B4}(d). 
These results suggest that analysis of meron and skyrmion helps understand the structure of the fractional vortices in various multi-component superconductors reported previously~\cite{RevModPhys.63.239,10.1143/PTP.102.965,PhysRevLett.89.067001,PhysRevLett.92.157001,PhysRevB.70.100502,PhysRevB.71.172510,TANAKA201844,PhysRevLett.119.167001,PhysRevResearch.2.043192,PhysRevB.83.144519,PhysRevB.86.024512,doi:10.1126/science.abp9979}.


Lastly, we study vortex flow across the nematic twin boundary, applying supercurrent $J_0$ parallel to the twin boundary by the gradient of the external field as $H=H_0 + J_0(-x+y)/\sqrt{2}\kappa^2$.  
Figure \ref{fig:B4w}(f) shows the vortex flow process when the conventional vortex is trapped in the twin boundary and changes to two fractional vortices, creating a skyrmion of vector ${\bm n}$ with $Q=-\frac{1}{2}-\frac{1}{2}$.
In Fig. \ref{fig:B4w}(g), two fractional vortices are combined into one conventional vortex when they escape from the twin boundary. 
We hope for future experimental observation of vortex motion~\cite{PhysRevLett.95.087002} to find the vortex flow as evidence of the fractional vortices.
It is noted that the vortex flow direction is tilted from the perpendicular direction to the twin boundary, reflecting anisotropy of the nematic vortex core, and the direction is changed across the twin boundary. 


In summary, we have investigated the spatial structure and the flow of vortex states near the nematic twin boundary in a nematic $s \pm d$ wave superconductor FeSe within a framework of two-component GL theory.  
There, orientation of the nematic vortex core shape changes $90^\circ$ from $s+d$ to $s-d$ wave pairing domains.   
We found vortices trapped at the nematic twin boundary may be fractional vortices with half-quantum, which have a topological nature of meron and skyrmion.   
This indicates that twin boundary of nematic superconductors such as FeSe may be a promising platform for research of fractional vortices in multi-component superconductors, which has been searched for a long time. 
To confirm the fractional vortices, it is preferable to search the vortex flow process in which the nematic vortex core is divided into two fractional vortices at the nematic twin boundary, as shown in our results, in addition to quantitative evaluation of magnetic flux per vortex~\cite{TANAKA201844,doi:10.1126/science.abp9979}. 
We expect that these experimental studies will be performed in the future.  
These studies also will be a test to examine whether $s \pm d$ wave GL theory can be an appropriate model to describe the vortex state and twin boundary state in a nematic FeSe superconductor.

\begin{acknowledgments}
This work is supported by JSPS KAKENHI Grant No. JP21K03471 and No. JP22H01941.
\end{acknowledgments}



%


\end{document}